\documentclass[aps,twocolumn,prb,amsmath,amssymb]{revtex4}
\usepackage{multirow}
\usepackage{graphicx,epsfig}
\usepackage{wasysym}

\begin{document}
\title{Thermoelectrical detection of Majorana states}
\author{Rosa L\'opez}
\affiliation{Institut de F\'{\i}sica Interdisciplin\`aria i de Sistemes Complexos
IFISC (CSIC-UIB), E-07122 Palma de Mallorca, Spain}
\affiliation{Kavli Institute for Theoretical Physics, University of California, Santa Barbara, California 93106-4030, USA}
\author{Minchul Lee}
\affiliation{Department of Applied Physics, College of Applied Science, Kyung Hee University, Yongin 446-701, Korea}
\author{Llorens Serra}
\affiliation{Institut de F\'{\i}sica Interdisciplin\`aria i de Sistemes Complexos
IFISC (CSIC-UIB), E-07122 Palma de Mallorca, Spain}
\affiliation{Departament de F\'{\i}sica, Universitat de les Illes Balears, E-07122 Palma de Mallorca, Spain}
\author{Jong Soo Lim}
\affiliation{School of Physics, Korea Institute for Advanced Study, Seoul 130-722, Korea}
\begin{abstract}
We discuss the thermoelectrical properties of nanowires hosting Majorana edge states. 
For a Majorana nanowire directly coupled to two normal reservoirs
the thermopower always vanishes regardeless of
the value of the Majorana hybridization. This situation changes drastically 
if we insert a quantum dot. Then,  the dot Majorana side coupled
system exhibits a different behavior for the thermopower depending on the 
Majorana hybridization parameter $\varepsilon_M$.
Thermopower reverses its sign when the half fermionic state is fully developed, 
i.e., when $\varepsilon_M=0$. As long as $\varepsilon_M$ becomes finite
the Seebeck coefficient behaves similarly to a resonant level system. The sign 
change of the thermopower when Majorana physics takes
place and the fact that both, the electrical and thermal conductances reach, 
their half fermionic value could serve as a proof of the existence
of Majorana edge states in nanowires. Finally, we perform some predictions 
about the gate dependence of the Seebeck coefficient
when Kondo correlations are present in the dot.
\end{abstract}
\maketitle

\section{Introduction}
Nowadays there is a lot of interest in the interplay
between heat and charge flows in nanostructures. \cite{Dhar08,Dubi11}
Thermovoltages generated in response to a temperature
gradient have been shown to be much bigger at the nanoscale  due
to the peculiar properties of quantum systems. \cite{Butcher90,Mabesoone92,Dzurak97,Godjin99,Matthews12}
For example,  delta like density of states occurring in confined 
nanostructures like quantum wells, \cite{Molenkamp92} alter 
dramatically their thermolectrical properties.
The main utility of thermoelectrical devices is the heat-to-
electricity conversion processes. However, from a more fundamental 
point of view, both thermal and electrical transport reveal information
on the intrinsic nature of a quantum system. An instance is the 
departure of the Wiedemann  Franz law  attributed to 
the non Fermi liquid  behavior. \cite{Coleman05}
In addition, thermoelectric transport measurements are able to distinguish between
 distinct types of carriers,  like electrons and holes in
  Andreev systems \cite{Jacquod10,Balachandran12} and molecular junctions. \cite{Reddy07}
 \begin{figure}[!h]
\centering
\includegraphics[width=0.4\textwidth]{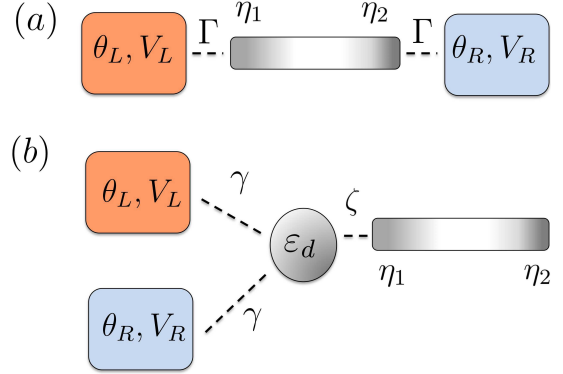}
\caption{(a) Majorana nanowire tunnel coupled to two normal contacts 
by tunneling barriers of probability $\Gamma$. Here, $\eta_1$, and $\eta_2$ denote the two
Majorana ends states at the semiconductor nanowire. Left(right) metallic 
contact is electrical and thermal biased with $V_{L(R)}$, and $\theta_{L(R)}$. (b) 
A quantum dot is inserted and symmetrically coupled to 
the metallic reservoirs with tunneling rate $\gamma$. 
The dot is side coupled to the Majorana nanowire, such coupling 
is characterized by the parameter $\zeta$.}
\label{figure1}
\end{figure}
 
 Our motivation is to address to what extent Majorana physics 
can be reflected in the thermoelectrical transport properties of a system.  
The unambiguous detection of Majorana fermions
 in solid state devices is still a discussional issue. Majorana physics,
  in the low energy domain, was predicted to occur as
 quasiparticle excitations. \cite{Wilczek09} The first proposals suggested
 their observation in quantum Hall states, 
 the Moore Read state at filling factor $\nu=5/2$. \cite{Read91}  Then, other suggestions considered
 some exotic superconductors like Sr$_2$RuO$_4$ 
 or $p$-wave superconductors. \cite{Kitaev01,Ivanov01,Sarma06,Linder10a}
Later on, the pioneering  work by Fu and Kane \cite{Ku08} demonstrated that such quasiparticles 
 could be created in a topological insulator brought in close proximity 
 with a superconductivity source.  However, the Majorana search 
 has been very prolific in the realm of quasi one dimensional 
 semiconductor nanowires \cite{Yuval10,Alicea10,Lutchyn10,Linder10b,Potter11},
and in particularly in large $g$ factor materials like InAs and InSb.  Most 
 of  the experiments designed to detect these elusive quasipartices have been performed via
electrical transport measurements \cite{Mourik12,Deng12,Heiblum12,Churchill12,Finck13} by tunnel
spectroscopy. A  voltage shift, $\delta V$, is applied to the nanowire edges that generates an electrical 
current $I$. The Majorana signature appears 
as a zero bias anomaly  in the nonlinear conductance $dI/dV$.\cite{Liu12,Pientka12,Elsa12a,Lim12}
In semiconductor nanowires, Majorana quasiparticles arise when
 superconductivity (source of electrons and holes), 
strong spin orbit interaction, and magnetic field work together. 
Then, under certain conditions the nanowire enters in the named
topological phase and shows up spinless, chargeless zero energy states, 
very elusive quasiparticle excitations. We refer to this as Majorana nanowire.
 However, the presence of a zero bias anomaly  in the nonlinear conductance
 does not warrant the presence of Majorana quasiparticles. 
 Kondo physics can be observed in normal superconductor nanowires  as well. \cite{Chang13,Eduardo12}
Furthermore, nearly zero energy Andreev states \cite{Kells12, Eduardo13} or 
 weak antilocalization \cite{Pikulin12a} effects are possible sources of zero bias anomaly in normal superconductor nanowires. 
 There are other suggestions  to detect Majorana zero energy states in 
Josephson junctions and rings. \cite{Kwon03,Fu09,Tanaka09,Ioselevich11,Jiang11,
Pikulin12b,Elsa12b,Fernando12} The Josephson current displays 
an anomalous periodicity of $4\pi$ if Majorana physics takes place. 
However, so far the experimental verification is not yet definitive. \cite{Rokhinson12}

Our goal consists in utilizing the thermoelectrical properties as a tool to detect 
the presence of Majorana edge states formed in normal superconductor nanowires. The only 
attempt to study similar
issues has done in $p$-wave superconductors. \cite{Refael13} Here,
we propose a way  of detecting Majorana edge states in semiconductor nanowires when 
a temperature gradient ($\delta\theta=\theta_L-\theta_R$) is applied and an induced electrical 
shift ($\delta V=V_L-V_R$) is generated.  We analyze a two terminal 
device as depicted in Fig. \ref{figure1}(a) and determine 
both the electrical and energy currents. Here,  the Majorana nanowire 
is contacted to two normal reservoirs. 
In general,  the linear response electric $I$ and energy $J$
currents can be expressed as
\begin{equation}\label{eq_matrix}
\begin{pmatrix}
I\\  J
\end{pmatrix}
=
\begin{pmatrix}
G& L\\ M& K
\end{pmatrix}
\begin{pmatrix}
\delta V\\ \delta \theta
\end{pmatrix} \,.
\end{equation}
The $2\times 2$ matrix is the Onsager matrix that includes diagonal
elements,the  electric $G$ and thermal $K$ conductances, and non diagonal coefficients, 
the thermoelectric $L$ and electrothermal $M$ conductances. The two latter are related 
due to microreversibility condition. \cite{Onsager31,Casimir45}
More specifically, we are interested in the determination of the Seebeck 
coefficient or thermopower that measures how efficient is the conversion 
of heat into electricity  in a thermoelectrical machine. The larger the Seebeck coefficient,
  the more efficient this conversion is. Seebeck coefficient is easily 
 determined from the relation: $S=-\delta V/\delta\theta=L/G$.  
 
Our results for a two terminal Majorana nanowire [see Fig. \ref{figure1}(a)] show that
both, the electrical and heat conductances reach their maximum 
value \textit{only} when Majorana edge states do not overlap. On
the contrary,  the thermoelectrical(electrothermal) 
response always vanishes irrespectively of the Majorana 
hybridization. 
As a result, the Seebeck coefficient vanishes owing to the
intrinsic particle hole symmetry of the system under consideration. 
However, this physical scenario can be
dramatically altered by inserting a quantum dot 
in between the two normal contacts and side coupled 
to the Majorana nanowire. \cite{Flensberg11,Dong11,Minchul12} Figure. \ref{figure1}(b)
 illustrates the sample configuration. In this arrangement, 
 the Seebeck coefficient can be 
tuned by gating the dot i.e., $S=S(\varepsilon_d)$
with $\varepsilon_d$ the dot level position.

\section{General formalism}
We present our theory for the thermoelectrical transport  by employing 
the nonequilibrium Keldysh Green function framework.  
We consider a semiconductor nanowire with strong Rashba spin orbit interaction
 with proximity induced $s$-wave superconductivity, and a applied magnetic
field $B$. We assume a sufficiently long wire to neglect 
charging effects.  The magnetic field is such that the wire is in the topological 
phase, $\Delta_Z>\sqrt{\Delta^2+\mu^2}$, with
$\Delta_Z=g\mu_B B/ 2$, and  $\mu$ the wire chemical 
potential. Then, isolated Majorana zero energy states $\eta_1=f+f\dagger$, 
and $\eta_2=i(f^\dagger-f)$ (in terms of $f$ Dirac fermions) 
are formed at the nanowire ends points. We consider that 
two normal contacts  are tunnel coupled to the wire ends 
as shown in Fig. \ref{figure1}(a). The 
Hamiltonian describing this system is given by  these three contributions:
$\mathcal{H} = \mathcal{H}_C + \mathcal{H}_M + \mathcal{H}_T$, where
\begin{eqnarray}\label{hamiltonian1}
\mathcal{H}_C &= &\sum_{\alpha,k} \varepsilon_{\alpha k} 
c_{\alpha k}^{\dagger} c_{\alpha k},\\ \nonumber
\mathcal{H}_M &=& \frac{i}{2} \varepsilon_M \eta_1 \eta_2 \,,\\ \nonumber
\mathcal{H}_T &=&\mathcal{H}_{TL}+\mathcal{H}_{TR}=\sum_{\alpha,k;\beta} \left[V_{\alpha k,\beta}
^{\ast} c_{\alpha k}^{\dagger} \eta_{\beta} + 
V_{\alpha k,\beta} \eta_{\beta} c_{\alpha k}\right] \nonumber\,.
\end{eqnarray}
Here, $\mathcal{H}_C$ describes the two normal leads, with 
$c^\dagger_{\alpha k}(c_{\alpha k})$  being the creation
(annihilation) operator for an electron with wavevector $k$ in the lead 
$\alpha$. Note that the spin degree of freedom 
is omitted. This can be understood considering that we need 
to apply a large magnetic field to observe the edge
Majoranas, so that only one kind of spin is effectively involved. 
$\mathcal{H}_{M}$ characterizes the
coupling between the two end Majorana states where
$\varepsilon_M \sim f(B,\Delta)e^{-L/\xi_0}$ with $L$  the length of the wire and $\xi_0$ the 
superconducting coherence length.  $f(B,\Delta)$ is a complicated function of $B$ and $\Delta$
that determines $\varepsilon_M$. 
 For our purpose we assume that $\varepsilon_M$ is a parameter. 
 The last contribution, $\mathcal{H}_T$ corresponds to the tunnel Hamiltonian 
between normal leads and the Majorana end states. Below, the tunnel amplitude $V_{\alpha k,\beta}$ 
is taken as $V_0$ for $\alpha=\beta$ and zero for $\alpha\neq \beta$. This defines
$\Gamma=\pi V_0^2 \rho_0$, with $\rho_0$ the contact density of states.

The charge and energy currents have 
the Landauer and B\"{u}ttiker form 
\begin{equation}
I=\frac{e}{h}\int d\omega \mathcal{T}(\omega)[f_{L}(\omega)-f_R(\omega)]\,,
\end{equation}
and 
\begin{equation}
J=\frac{1}{h}\int d\omega \omega\mathcal{T}(\omega)[f_{L}(\omega)-f_R(\omega)]\,,
\end{equation}
with a transmission coefficient given by
\begin{equation}
\mathcal{T}(\omega)=\frac{4\Gamma^2\left(\omega^2
+4\Gamma^2+\varepsilon_M^2\right)}{\left(\omega^2
+4\Gamma^2\right)^2+\varepsilon_M^2\left[\varepsilon_M^2
-2\left(\omega^2-4\Gamma^2\right)\right]}\,.
\end{equation}

Here $f_L=1/[1+\exp{(\omega-(\mu+V_L))/k_B\theta_L}+1]$ 
($k_B$ Boltzamnn constant) and 
$f_R=1/[1+\exp{(\omega-(\mu+V_R)/k_B\theta_R}+1]$ are 
the Fermi Dirac distribution
function for the left and right contacts respectively with $V_{L,R}=\pm\delta V/2$,  and 
$\theta_{L,R}=T_b\pm\delta \theta/2$. 

%
The linear conductances are  (we take $\mu=0$)
\begin{eqnarray}
G&=&\frac{e^2}{h}\int d\omega\mathcal{T}(\omega) \left[-\frac{\partial f_{eq}}{\partial \omega}\right], \\ 
L&=&\frac{e}{hT_b}\int d\omega \omega \mathcal{T}(\omega) \left[-\frac{\partial f_{eq}}{\partial \varepsilon}\right], \\ 
M&=&\frac{e}{h}\int d\omega \omega\mathcal{T}(\omega) \left[-\frac{\partial f_{eq}}{\partial \omega}\right], \\ 
K&=&\frac{1}{hT_b}\int d\omega \omega^2 \mathcal{T}(\omega) \left[-\frac{\partial f_{eq}}{\partial \omega}\right] ,
\end{eqnarray}
where $f_{eq}$ is the equilibrium Fermi Dirac distribution function when $\delta T=0$ and $\delta V=0$.
In a Sommerfeld expansion, at sufficiently low temperatures,
the linear response conductances $G$, and $K$
 have the same behavior with the transmission coefficient up to a proportionality factor: $G_0$, and $K_0$. Thus, 
\begin{equation}
G(K)\! =\!\lim_{\delta V\rightarrow 0} \frac{dI}{dV}	\left(\lim_{\delta\theta\rightarrow 0}\frac{dJ}{d\theta}\right)\!\!=\!\!
G_0(K_0)\frac{4\Gamma^2}{\varepsilon_M^2+4\Gamma^2}.
\label{eq:LinearG0}
\end{equation}
 with $G_0=e^2/h$ (quantum electrical conductance), and $K_0= \pi^2 k_B^2 T_b/3h$ (quantum thermal conductance).
 They take their maximum value $G_0$, and $K_0$, respectively when $\varepsilon_M=0$, otherwise, they vanish as $\varepsilon_M$ grows.
Importantly, the
 off diagonal conductances are always zero,  $L=L_0\partial \mathcal{T}(\omega)/\partial\omega|_{\omega=0}$ 
 with $L_0=e\pi^2 k_B^2 T_b/3h$  (and $M=L /T_b$).  The vanishing value of the $L(M)$ has profound consequences in the thermopower or Seebeck coefficient
(we recall that $S=L/G$). The Seebeck coefficient vanishes regardless of the value of $\varepsilon_M$. 
The  reason for this result lies
in the inherent particle hole symmetry of our system, there is no electrical response to a thermal gradient. 

Asymmetry in the particle and hole subspaces can happen if we 
insert a quantum dot between the two normal contacts. Here the dot is side coupled to the 
Majorana as illustrated in Fig. \ref{figure1}(b).  The thermoelectrical transport through the 
dot Majorana system shows a non zero value for
the off diagonal Onsager conductances when the dot is off resonance, 
i.e.,  a nonzero Seebeck coefficient. 
Importantly, we can tune the Seebeck coefficient from  zero 
when the dot is on resonance 
to large values when is off resonance. Besides, the behavior of the 
Seebeck coefficient with the dot level is quite different depending 
on the value of the Majorana hybridization parameter, $\varepsilon_M$. 
Thus, Seebeck coefficient might allow us to detect truly zero energy Majorana states
for which $\varepsilon_M$ is negligible .

\section{Side tunel coupled dot Majorana system}
In order to include the quantum dot we need to reformulate the Hamiltonian as follows. 
First, we consider the dot Hamiltonian 
\begin{equation}
\mathcal{H}_d=\sum\varepsilon_d d^\dagger d\,,
\end{equation}
where $d(d^\dagger)$ operator annihilates(creates) an 
electron  on the dot site.  We consider a single dot level with energy $\varepsilon_d$. 
The dot is  connected to the left and right normal contacts by tunnel barriers
\begin{equation}
\mathcal{H}_{Td}=\sum_{\alpha k} (W_{\alpha } c^\dagger_{\alpha k} d + h.c)\,.
\end{equation}
We consider symmetrically dot 
coupling to the normal contacts with a common tunneling rate: $\gamma=\pi W^2\rho_0$, with $W=W_{L}=W_{R}$.
The dot is side coupled to the Majorana nanowire as
\begin{equation} 
 \mathcal{H}_{TM}=\sum_{\beta} \zeta (d^\dagger \eta_\beta+\eta_\beta d)\,,
\end{equation} 
with $\beta=1,2$. Here, we assume that only the closest Majorana state to the dot is coupled, say $\eta_1$.   
The total Hamiltonian is the sum of all these contributions, and the contact and 
Majorana Hamiltonians [$\mathcal{H}_C$, and $\mathcal{H}_M$, see Eq.  (\ref{hamiltonian1})]: $\mathcal{H}=\mathcal{H}_C+\mathcal{H}_d+\mathcal{H}_{M}+\mathcal{H}_{Td}+ \mathcal{H}_{TM}$.
%
Now, the charge and energy flows can be expressed in terms of the dot transmission (see Ref. [\onlinecite{Dong11}] for details)
\begin{equation}\label{transmission}
\mathcal{T}_d(\omega)=-\frac{1}{2}\frac{\gamma}{\pi} \rm{Im} \mathcal{G}^r_{d}(\omega) \,,
\end{equation}
where $\mathcal{G}^r_{d}$ is the retarded dot Green function
\begin{equation}
\mathcal{G}^r_{d}(\omega)=\frac{1}{\omega-\varepsilon_{d}+i\frac{\gamma}{2} - B(\omega)\left[1+\tilde{B}(\omega)\right]}\,,
\end{equation}
with
\begin{eqnarray}
\tilde{B}(\omega)=\frac{B(\omega)}{\omega+\varepsilon_{d}+i\frac{\gamma}{2}- B(\omega)}.
\end{eqnarray}
The parameter $\zeta$  in Eq. (\ref{transmission}) characterizes the dot Majorana coupling 
where $B(\omega)=|\zeta|^2/(\omega-\varepsilon_M^2/\omega)$ 
being the dot Majorana selfenergy coupling.


\section{Discussion}
\begin{figure}
 \centering
    \includegraphics[width=0.45\textwidth]{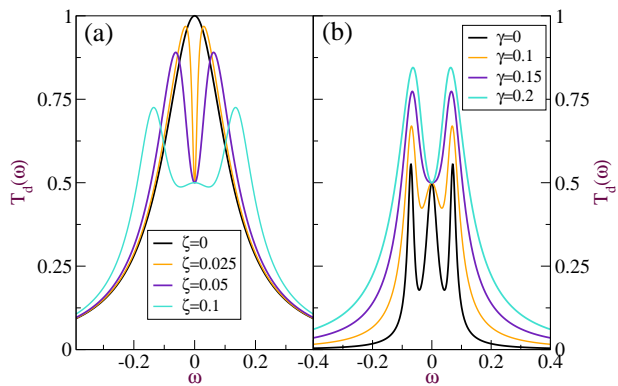}
  \caption{Dot transmission $\mathcal{T}_d(\omega)$  for (a) various $\zeta$ values as
   indicated and $\gamma=0.25$;  (b) for different $\gamma$ values and $\zeta=0.05$. 
   Parameters: $\varepsilon_d=0.0$, $\varepsilon_M=0$.} 
\label{figure2}
\end{figure}

Before starting the discussion of the thermoelectrical properties
in the dot Majorana system it is worth to revisit the behavior of the
dot transmission with the system parameters, $\varepsilon_M$, $\varepsilon_d$,
$\zeta$ and $\gamma$. \cite{Dong11} Hereafter, we employ 
$D=50$ for the contact bandwidth 
that determines our energy unit. The dependence of $\mathcal{T}_d(\omega)$ 
with $\zeta$, and $\gamma$  is illustrated in Fig. \ref{figure2}
 when the dot is on resonance 
and no Majorana overlap occurs ($\varepsilon_d=0$, and  $\varepsilon_M=0$). 
For the uncoupled Majorana situation
the transmission corresponds to the resonant level model 
with unitary transmission. As $\zeta$ is turn on
two peaks at $\omega=\pm\zeta$ appear 
due to the dot Majorana finite coupling. Now, keeping 
 fixed $\zeta$ and tuning $\gamma$ the dot transmission shows 
  a three peak structure when $\gamma\approx \zeta$ in which the zero energy peak
   is the signature of the presence of Majorana edge states [see Fig. \ref{figure2}(b)] . 
  In all cases, when $\zeta\neq 0$, the 
dot transmission is always half fermionic. \cite{Dong11,Minchul12}

\begin{figure}
  \centering
    \includegraphics[width=0.45\textwidth]{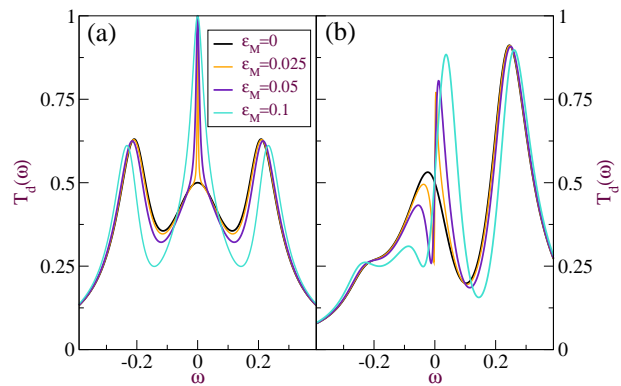}
  \caption{Dot transmission, $\mathcal{T}_d(\omega)$ for different 
  values of the Majorana overlap $\varepsilon_M$ (a) for 
  $\varepsilon_d=0$, and (b) for $\varepsilon_d=0.12$. 
Parameters:   $\gamma=0.25$, $\zeta=0.15$.}
\label{figure3}
\end{figure}

When $\varepsilon_M$ acquires a finite value, $\mathcal{T}_d$ becomes unitary, as shown in 
Fig. \ref{figure3}(a). For large $\varepsilon_M$,  $\mathcal{T}_d$ corresponds to
the one for a resonant level mode,  with resonances at $\omega\pm\epsilon_M$ 
due to the coupling of the dot state with the $f$ Dirac fermions in the wire 
(resulting from the large  Majorana hybridization).

Thermoelectrical effects appears when the transmission becomes asymmetric. In order to observe such 
asymmetric transmission for $\omega<0$, and $\omega>0$
the dot level must be positioned off resonance, i.e., $\varepsilon_d\neq 0$. This situation is presented in Fig. \ref{figure3}(b) for several
values of the Majorana hybridization parameter when $\varepsilon_d=0.12$. Note that, the transmission is asymmetric  even for $\varepsilon_M=0$
although is still half fermionic.  For a nonzero Majorana overlap, the transmission depends strongly on the dot gate value 
leading to a non unitary electrical(thermal) conductance. 

\begin{figure}
  \centering
    \includegraphics[width=0.45\textwidth]{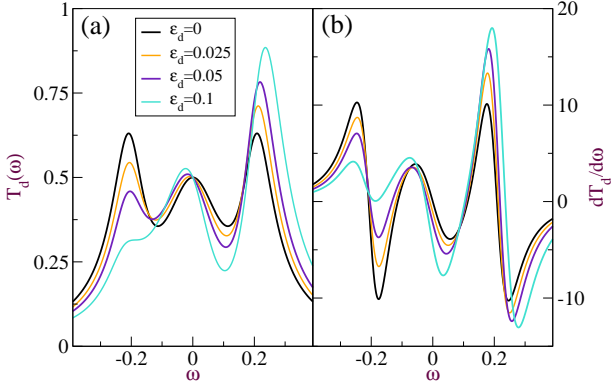}
  \caption{ (a) Dot transmission $\mathcal{A}_d(\omega)$  and 
  (b) its derivative $\partial \mathcal{T}_d(\omega)/\partial \omega$ for the
 indicated $\varepsilon_d$ values and $\varepsilon_M=0$. Parameters:  $\gamma=0.25$, $\zeta=0.15$. }
\label{figure4}
\end{figure}

The  dot gate dependence of $\mathcal{T}_d(\omega)$ for an ideal Majorana nanowire ($\varepsilon_M=0$)
is depicted in Fig. \ref{figure4}(a) and its energy derivative in Fig. \ref{figure4}(b). These curves shown
that the transmission at zero energy is always half fermionic as should be for $\varepsilon_M=0$, regardless
of the dot gate value. However, it is interesting to observe that the energy derivative of the transmission at zero energy 
acquires some dot gate dependence reflecting the asymmetry between the particle and hole sectors. 
This result is important for the thermoelectrical conductance $L$,we recall that $L=L_0\partial T_b(\omega)/\partial\omega|_{\omega=0}$ implying that
$L$ becomes gate dependent. Whereas the diagonal conductances are not sensitive to the particle hole asymmetry
introduced by $\varepsilon_d\neq 0$, the off diagonal conductances 
show a dot gate dependence with important consequences in the thermoelectrical
transport.

\begin{figure}
\centering
    \includegraphics[width=0.45\textwidth]{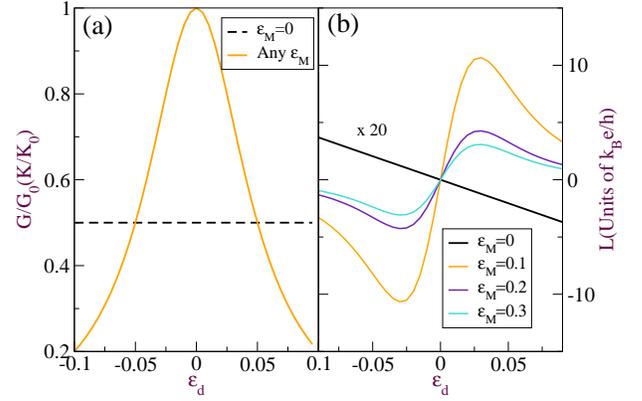}
  \caption{ (a) Dot gate dependence of the linear electrical(thermal) conductance $G(K)$ (with $G_0=e^2/h$, $K_0=\pi^2k_B^2 T_b/3h$)
   for zero $\varepsilon_M=0$ and 
  finite Majorana overlap $\varepsilon_M\neq 0$. (b) Thermoelectrical conductance $L$ versus
  $\varepsilon_d$ at different Majorana overlaps $\varepsilon_M$. The case $\varepsilon_M=0$ has been 
  multiplied by a factor $20$ for comparison purposes. Parameters: $\gamma=0.25$, $\zeta=0.15$, and $T_b=0.025$.}
\label{figure5}
\end{figure}

Our previous analysis for the dot transmission 
explains the curves for  the 
conductances illustrated in Fig. \ref{figure5}. Both,  the  electrical and thermal conductances,
$G$, and $K$ depend strongly on $\varepsilon_d$ whenever 
the two end Majorana states overlap. Otherwise, in the ideal situation where
$\varepsilon_M=0$, $G$, and $K$ take its maximum value and they becomes half fermionic. \cite{Beenakker11,Dong11,Minchul12} 
This important result it serves to us to detect
the presence of Majorana edge states in side coupled dot nanowires systems. However, the previous results
are applicable only for purely electrical or thermal transport measurements. Here, we 
are interested more in the thermolectrical signatures of the Majorana edge states. For that purpose,
 we analyze how the off diagonal conductances behave with the dot gate values. We find, that  
when Majorana edge states  have negligible overlap ( i.e., $\varepsilon_M=0$)
the off diagonal conductance $L(M)$ reverses it sign in 
comparison with a situation with finite overlap, i.e.,  $\varepsilon_M\neq 0$. Our results show
that for zero Majorana overlap $\varepsilon_M=0$, the thermoelectrical conductance $L$ depends linearly 
with $\varepsilon_d $ with a negative slope $-1/2\zeta^2$ that depends inversely on the dot Majorana strength. 
However, for a finite  Majorana overlap, when $\varepsilon_M\neq 0$ the thermoelectrical conductance
  $L/L_0=[\varepsilon_d/ (4\varepsilon_d^2+\gamma^2)^2][8\gamma^2(\varepsilon_M^2+\zeta^2)/\varepsilon_M^2] $,  displays
  two extrema at $\varepsilon_d=\pm\gamma/2$. In this case, $L$ behaves similarly to the resonant level model. 
  Importantly, the different behavior found
for the gate dependence of the thermoelectrical conductance $L$ could be utilized as an
\textit{smooking gun} for the Majorana detection in thermoelectrical transport measurements. 

\begin{figure}
  \centering
    \includegraphics[width=0.45\textwidth]{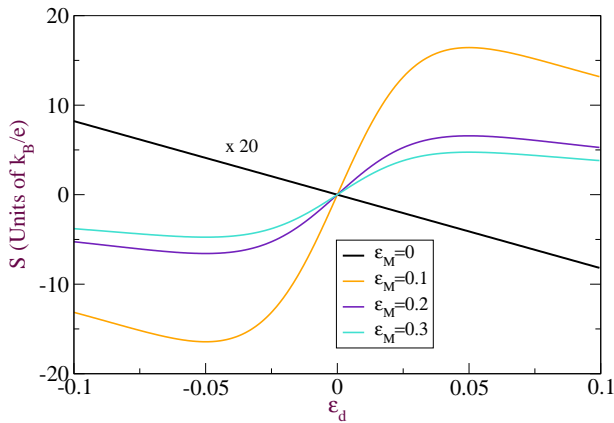}
  \caption{Thermopower $S$ versus $\varepsilon_d$  
  for various $\varepsilon_M$ values. The curve 
  corresponding to $\varepsilon_M=0$ has been 
  enlarged by a factor $20$ for comparison purposes.  Parameters: $\gamma=0.25$, $\zeta=0.15$, and $T_b=0.025$.}
      \label{figure6}
\end{figure}

Using the previous results, we discuss the gate dependence of the thermopower 
$S=L/G=-\delta V/\delta \theta$, where $S=(\pi^2 k_B^2T_b/3e) d\ln\mathcal{T}(\omega)/d\omega|_{\omega=0}$ is the Mott
 formula. We define $S_0=\pi^2 k_B^2T_b/3e$. For the dot Majorana uncoupled case, $\zeta=0$,  the thermopower $S/S_0=8\varepsilon_d/(4\varepsilon_d^2+\gamma^2)$
 vanishes when $\varepsilon_d=0$ and follows the resonant level model as expected. For the coupled system, when $\zeta\neq 0$, the thermopower
$S$ versus the dot gate position is plotted in Fig. \ref{figure6}. Remarkably,  the thermopower  is linear with $\varepsilon_d$ for zero
Majorana overlap: $S/S_0=-\varepsilon_d/\zeta^2$ when $\varepsilon_M=0$ and $\zeta\neq 0$. 
The dot gate dependence of $S$ is due to the particle hole asymmetry
introduced when  $\varepsilon_d$ is tuned from the \textit{on} to the \textit{off} resonance situation. 
 The way to understand this result  is by the addition of two effects. First, the Majorana state
contributes to the thermopower in a rigid way with a constant term $-1/\zeta^2$. Second, the particle hole asymmetry grows as $\varepsilon_d$ does and this explains why the thermopower 
 grows with $\varepsilon_d$. Then, both 
features add up and produce a linear dependence of the Seebeck coefficient with the dot 
gate with a negative slope that depends on the inverse of the dot Majorana coupling $\zeta$. 

Figure \ref{figure6} displays our results for the thermopower for various values of $\varepsilon_M$.
For $\varepsilon_M=0$,  Fig. \ref{figure6} shows that the thermopower is positive(negative)  
for negative(positive) $\varepsilon_d$ having $\delta V<0$ by heating up(cooling down) 
the left contact. The thermopower sign dependence with $\varepsilon_d$ is inverted  when the Majorana overlap is finite. 
Here, for $\varepsilon_M$ finite the thermopower is: $S/S_0= [\varepsilon_d/(4\gamma^2+\varepsilon_d^2)][8(\varepsilon_M^2+\zeta^2)/\varepsilon_M^2)]$.  
This means that when $\varepsilon_d<0(\varepsilon_d>0)$  the heating(cooling)
of the left contact  induced a positive(negative) voltage difference. Here, 
 the Seebeck coefficient follows the behavior for a resonant model with two extrema
 at $\varepsilon_d=\pm \gamma/2$.   
 All these differences for $S(\varepsilon_d)$ depending on $\varepsilon_M$
it allows us to distinguish situations where nanowires can host truly Majorana edge states or not.

Some of the previous results allow us to predict the  dot gate dependence of the Seebeck coefficient, when Coulomb 
interactions take place. A quantum dot with a free local moment is able to form a Kondo singlet with the
delocalized electrons in the normal reservoirs when is strongly tunnel coupled to them. 
Then, at temperatures much lower than the Kondo scale $T_K$ the dot physics can be explained 
 within the Fermi Liquid theory. \cite{sbmft} In this scenario, both the dot gate position
$\tilde{\varepsilon}_d\rightarrow \varepsilon_d+\lambda$, and the lead dot tunneling rate $\Gamma\rightarrow \tilde{\Gamma}$
are renormalized by  Kondo correlations as $\lambda=-\varepsilon_d$, and $\tilde{\Gamma}=T_K$. Under this situation, the Seebeck
coefficient, in the Kondo regime is zero (with $T_K$ larger that the dot Majorana coupling 
selfenergy \cite{Minchul12}, i.e., in the Kondo dominant
regime). In the pure Kondo regime spin fluctuations carry the charge and energy transport in a particle and hole symmetric situation,
then, it quite reasonable to expect a vanishing Seebeck coefficient no matter the Majorana overlap is.  For more exotic Kondo effects 
in which particle hole symmetry breaks down, like in the SU(4) Kondo effect (recently observed in carbon nanotube quantum dots \cite{jarillo05,rosa05}) 
a nonvanishing Seebeck effect is expected. Here, within the Fermi Liquid
description we have $\tilde{\varepsilon}_d\approx T_K^{SU(4)}$, and $\tilde{\Gamma}=T_K^{SU(4)}$ [with $T_K^{SU(4)}$ as the Kondo scale for the SU(4) case].
These two renormalized parameters produce a nonzero, but constant  Seebeck coefficients: $S(\varepsilon_d)\approx -T_K^{SU(4)}/\zeta^2$ 
when $\varepsilon_M=0$ and $S(\varepsilon_d)= c/T_K^{SU(4)}$ ($c>0$) when $\varepsilon_M$ is finite. 
The richness of the Kondo behavior when Majorana physics occurs has been
detailed discussed in Ref. [\onlinecite{Minchul12}] by some of the authors but only for the electrical transport. The understanding of the thermoelectrical properties
for the different range of parameters, i.e., in the Kondo and Majorana dominant regimes,  requires further analysis with more powerful
theoretical techniques \cite{progress}).
\section{Conclusion}
We have investigated the linear response conductances to a thermal and electrical voltage shift
in a two terminal geometry with normal superconductor nanowires showing Majorana physics. Firstly, we have considered a nanowire directly 
coupled to two normal reservoirs. Due to the intrinsic particle hole symmetry this system
exhibits a null thermopower, no voltage is generated in response to a thermal gradient. 
Then, we insert a quantum dot between the two normal contacts which
is side coupled to the Majorana nanowire. With this arrangement the detection of the Majorana edge states
can be performed by looking at the sign of the thermoelectrical conductance or the thermopower $S$.
Besides, we show that both, the electrical and thermal conductances take their half fermionic values
whenever a true Majorana fermion state is formed, when $\varepsilon_M=0$. Finally, we make
some predictions for the gate dependence of the Seebeck coefficient for interacting dots in the Kondo regime. 
We believe that our results
could serve as an unambiguous tool for the detection of Majorana edge states in semiconductor nanowires. 

\emph{Note added---}During the completion of this paper we become aware of a related
work dealing with thermolectric transport in normal-dot-Majorana nanowires systems.
The difference is that we consider thermal and electrical bias
applied to the normal contacts, in Ref. [\onlinecite{Martin13}] the thermoelectrical
forces are applied to the normal and Majorana parts. 
\section{Acknowledgement} 
We thank David S\'{a}nchez for useful discussions. Work supported by MINECO Grant No. FIS2011-23526.   
This research was supported in part by the
Kavli Institute for Theoretical Physics through NSF grant PHY11-25915.

\end{document}